\documentclass[draftcls, 12pt, onecolumn]{IEEEtran}
 \makeatletter

\newcommand{\Rmnum}[1]{\expandafter\@slowromancap\romannumeral #1@}
\makeatother
\usepackage{graphicx,cite,epsfig,amssymb,amsmath,multirow}
\usepackage{graphicx,amsmath,amssymb,amsfonts,cite}
\usepackage{mathrsfs}
\usepackage{algorithm}
\usepackage{algorithmic}
\usepackage{float}
\usepackage{ulem}
\usepackage{color}

\usepackage{stfloats}
\usepackage{setspace}
\makeatother
\usepackage{fancyhdr}

\usepackage{xcolor}
\usepackage{colortbl,booktabs}
\usepackage{ifpdf}
\usepackage{cite}
\usepackage{amsmath}
\usepackage{bm}
\usepackage{array}
\begin{document}
\bibliographystyle{ieeetr}


\title{MMSE Channel Estimation  for Two-Port
Demodulation Reference Signals in  New Radio}

\author{Dejin Kong, Xiang-Gen Xia, \IEEEmembership{Fellow, IEEE}, Pei Liu, and Qibiao Zhu
	
\thanks{D. Kong  is with the School of Electronic and Electrical Engineering, Wuhan Textile University, Wuhan, 430074, China (e-mail: djkou@wtu.edu.cn).}

\thanks{X.-G. Xia with the Department of Electrical and Computer Engineering,
University of Delaware, Newark, DE 19716, USA (e-mail: xxia@ee.udel.edu)}

\thanks{P. Liu (Corresponding author) is with the School of Information Engineering, Wuhan University of Technology, Wuhan 430070, China (e-mail: pei.liu@ieee.org).}

\thanks{Q. Zhu is with the School of Information Engineering, Nanchang University, Nanchang {\rm 330031}, China (e-mail: zhuqibiao@ncu.edu.cn).}

}


\maketitle

\begin{abstract}
Two-port demodulation reference signals (DMRS) have been employed in new radio (NR) recently.
In this paper, we firstly propose a minimum mean square error (MMSE) scheme
with full priori knowledge (F-MMSE) to achieve the channel estimation of two-port DMRS in NR. When the two ports are assigned to different users, the full priori knowledge of two ports is not easy to be obtained for one user. Then, we present a MMSE scheme
with partial priori knowledge (P-MMSE). Finally, numerical results show that the proposed schemes achieve satisfactory channel estimation performance.
Moreover, for both mean square error and bit error ratio metrics, the proposed schemes can achieve better performance compared with
the classical discrete Fourier transform based channel estimation. Particularly,  P-MMSE scheme delivers almost the same performance compared with F-MMSE scheme by a small amount of prior knowledge.
\end{abstract}


\IEEEpeerreviewmaketitle
\section{Introduction}
As the most well-known multicarrier modulation, orthogonal frequency division multiplexing
(OFDM) has been adopted in various wireless communication standards \cite{ref0,ref00,ref1,ref2.0}, due to the ease of implementation
and the ability to fight against multi-path fading channels \cite{ref2,ref2.11}.
Recently, it has been agreed by the third  generation partnership project (3GPP) that,
OFDM is employed as the physical-layer technique for both of the downlink and uplink in new radio (NR) \cite{ref3,ref3.0}
for the fifth-generation mobile communications (5G), and
demodulation reference signal (DMRS) with multiple ports are designed to
achieve the channel estimation as well as the equalization on data symbols.

\textcolor{black}{In NR, two types of DRMS are adopted, i.e., Type 1 and Type 2 \cite{ref3}. In Type 1,
a pair of adjacent subcarriers distinguish two DMRS ports by employing  the orthogonal cover code (OCC) and 2 of each 6 subcarriers are selected for the two DMRS ports. In Type 2,  half of the
frequency subcarriers are
selected and  cyclic shift phases are adopted to distinguish two DMRS ports.  In this paper, we
only focus on the channel estimation based on DMRS of Type 2.    }
In general, the channel estimation can be achieved by the  classical discrete Fourier transform
(DFT) scheme  \cite{ref4,ref5,ref6} since  channels for the two ports
of DMRS can be separated in the time domain by the DFT
operation.
Since the two DMRS ports in NR are equivalent to orthogonal by OCC,
the OCC-based method in \cite{Sesia} can be employed to separate the two
DMRS ports, based on the assumption that channel frequency responses at adjacent subcarriers are equal.
In \cite{ye}, a robust channel estimator was presented for OFDM systems and it was
shown that the minimum mean square error (MMSE) estimator is insensitive to the channel statistics.
\textcolor{black}{
Furthermore, in \cite{addzhang}, it was demonstrated that, the MMSE estimator is
robust to correlation matrix mismatch for the multi-user
scenario, which makes
the MMSE channel estimation  more practical.}
To the best of our knowledge, little is known about utilizing the MMSE metric
for the channel estimation of two-port DMRS in NR in the open literature.

In this paper, we propose a new approach for the two-port channel estimation in NR,
based on the MMSE metric.
Firstly, we propose an MMSE with full priori knowledge (F-MMSE) scheme to achieve the channel estimation of
two-port DMRS in NR.
Then, we present an MMSE
with partial priori knowledge (P-MMSE) scheme when
the two ports are assigned to different users and the full priori knowledge of two ports
is  not easy to obtain for one user.
Finally, theoretical analysis and numerical simulations are carried out to
validate the performance of the proposed schemes. They show that the proposed two schemes can achieve high quality channel estimation. They also show that the proposed two schemes perform far better than the  classical DFT scheme. More importantly, the P-MMSE scheme approaches to the same performance of the F-MMSE scheme with a little priori knowledge.

The remainder of this paper is organized as follows.  In Section II, the system model and
two-port of DMRS in NR are  presented briefly, as well as the classical DFT-based channel
estimation method and. In Section III,
the two-port MMSE channel estimation methods are proposed.
Simulation results are given in Section IV and this paper is concluded in Section V.

\textit {Notation:}  Lower-case  and upper-case  boldface letters denote vectors and matrices, respectively. ${\bf I}_{P}$ denotes a $P\times P$ identity matrix. $(\cdot)^{\text \dag}$, $(\cdot)^{\text T}$, and $(\cdot)^{-1}$ denote the Hermitian transpose, transpose, and inverse operations, respectively.  The expectation operation is $\mathbb{E}\{\cdot\}$. ${\text {diag}}\left({\bf a}\right)$ denotes a diagonal matrix where the main diagonal entries are the elements of vector ${\bf a}$. Finally, $\|\cdot\|_{2}$ denotes the 2-norm of a vector.

\section{Conventional Channel Estimation Schemes}

\subsection{System Model}

\begin{figure*}[!t]
	\centering
	\includegraphics[scale=0.6]{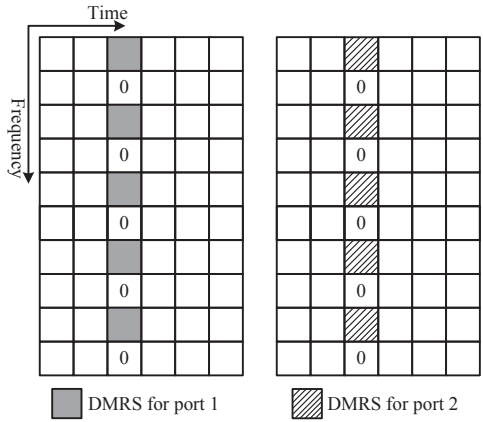}
	\caption{Frame structure of two DMRS ports in NR}
	\label{fig1}
\end{figure*}

We consider an OFDM system with $M$ subcarriers, equipped with two transmitting antennas and two receiving antennas.
Assume that $2P<M$  subcarriers are assigned to one user, two ports of DMRS are as shown in Figure~\ref{fig1}, which share
the same time frequency resources. Assume $x_p = a_{2p,q}$ is the pilot of DMRS with the transmitting power of 1, where $p=0,1,\ldots,P-1$ is  the frequency index and $q$ is the  time index,  Moreover,  $a_{2p+1,q}$ is usually set to zero unless it is used to support more DMRS ports. Hence, in this paper, $a_{2p+1,q}$ is set to zero. If $x_p$ is the pilot symbol of the first DMRS port, the pilot symbol of the second DMRS port would be $x_p e^{j2 \pi \frac{p\Delta_{\text {cs}}}{P}}$ \cite{ref3}, where $\Delta_{cs}$ is the cyclic
shift and it is  equal $\frac{P}{2}$ for the orthogonality of two DMRS ports.

Then, the demodulation of the $r$th receiving antenna can be written as
\begin{equation}\label{eq 2.1}
	y_p^{r}= h_{p}^{1r} x_p + h_{p}^{2r} x_p e^{j2 \pi \frac{p\Delta_{\text {cs}}}{P}} + \eta_p^{r},
\end{equation}
where $r=0$ or 1, $h_{p}^{tr}$ is the channel frequency response for $p$th pilot, between the $t$th transmitting antenna
and the $r$th receiving antenna, and $\eta_p^{r}$ is the additive white Gaussian noise
with mean zero and variance $\sigma^2$ \cite{ref9.11,ref9.1}.

For simplicity, the matrix form of (\ref{eq 2.1}) is obtained as
\begin{equation}\label{eq 2.2}
	\textbf{y}_{r}= \textbf{X}\textbf{h}_{1r} +\textbf{X} \textbf{C} \textbf{h}_{2r} + \boldsymbol{\eta}_{r},
\end{equation}
where $\textbf{y}_{r} = [y_0^{r},y_1^{r},\ldots,y_{P-1}^{r}]^{\text T}$, $\textbf{h}_{tr} = [h_{0}^{tr}, h_{1}^{tr},\ldots,h_{P-1}^{tr}]^{\text T}$, $\boldsymbol{\eta}_{r} = [\eta_0^{r}, \eta_1^{r}, \ldots,\eta_{P-1}^{r}]^{\text T}$,
and $\textbf{X}=\\{\text {diag}}(x_0,\ldots, x_p,\ldots,x_{P-1})$. Also, $\textbf{C}$ is a diagonal matrix
and its $p$th
diagonal element is $e^{j2 \pi \frac{p\Delta_{\text {cs}}}{P}}$.

Note that, $\textbf{X}$ is a diagonal and reversible matrix. Then, the two-port DMRS-based channel estimation model can be obtained \cite{ref6}, as
\begin{align}\label{eq 2.3}
\hat{\textbf{h}} = \textbf{X}^{-1}\textbf{y}_{r}
= \textbf{h}_{1r} + \textbf{C} \textbf{h}_{2r} + \textbf{X}^{-1}\boldsymbol{\eta}_{r},
\end{align}
where $\textbf{X}^{-1}$ is the inverse matrix of $\textbf{X}$.

In this paper, we assume that the two DMRS ports are assigned to different users. For downlink transmission, each user
has to perform channel estimation  independently.

\subsection{DFT-based Channel Estimation}

In this subsection, the DFT-based scheme is briefly presented to achieve the channel estimation of two-port DMRS.
Firstly, a $P$-point inverse DFT (IDFT) operation is performed on $\hat{\textbf{h}}$, and it is obtained as
\begin{align}\label{eq 2.4}
\hat{\underline{\textbf{h}}} = \textbf{F} \hat{\textbf{h}}
=\underline{{\textbf{h}}}_{1r} +\underline{{\textbf{h}}}_{2r}\left(\frac{P}{2}\right)  + \boldsymbol{\xi}_{r},
\end{align}
where $\textbf{F}$ is the IDFT matrix with dimension of $P \times P$, whose element at the position $(p,q)$ is $\frac{1}{\sqrt{P}}e^{j2\pi \frac{p q}{P}}$, and $\boldsymbol{\xi}_{r} =  \textbf{F}\textbf{X}^{-1}\boldsymbol{\eta}_{r}$. Also,
$\underline{{\textbf{h}}}_{tr} $ is the channel impulse response between the $t$th transmitting antenna
and the $r$th receiving antenna, and $\underline{{\textbf{h}}}_{2r}\left(\frac{P}{2}\right)$ is a cyclic shift
vector of $\underline{\textbf{h}}_{2r}$ with a shift $\frac{P}{2}$.
Note that, the number of subcarriers in OFDM should be at least twice larger than the channel length.

\begin{figure}[!t]
	\centering
	\includegraphics[scale=0.8]{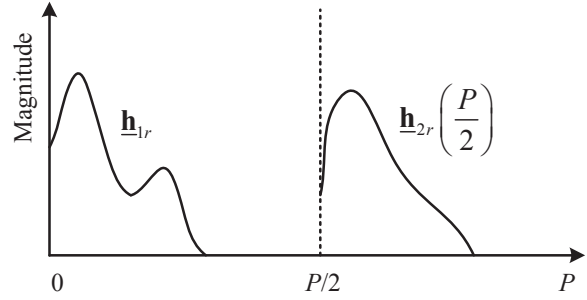}
	\caption{Separation of channels in the DFT-based channel estimation.}
	\label{fig2}
\end{figure}

From (\ref{eq 2.4}), it can be observed that, the channel impulse responses $\underline{\textbf{h}}_{1r}$
and $\underline{\textbf{h}}_{2r}$
are separated by $\frac{P}{2}$ in the time domain as shown in Figure~\ref{fig2}, and the time-domain channel estimation
can be performed by the following function,
\begin{align}\label{eq 2.5}
\begin{split}
   g(p)=\left\{
\begin{array}{*{20}l}
   1, &0\leq p<\frac{P}{2}-1,\\
   \\
   0, &\frac{P}{2}\leq p< P-1,\\
\end{array}\right.
\end{split}
\end{align}
Denote a $P \times P$ diagonal matrix $\textbf{G}$ with the  diagonal element $g(p)$.
Then, the channel estimation of $\underline{{\textbf{h}}}_{1r}$ can be obtained as
\begin{align}\label{eq 2.6}
\hat{\underline{{\textbf{h}}}}_{1r} = \textbf{G} \hat{\underline{{\textbf{h}}}}.
\end{align}
Then, the channel frequency responses can be obtained by a $P$-point DFT operation
\begin{align}\label{eq 2.7}
\hat{\textbf{h}}_{1r}^{\textrm{DFT}} = \textbf{G} \hat{\underline{{\textbf{h}}}}_{1r}.
\end{align}
As for ${\textbf{h}}_{2r}$, the channel estimation can be performed by the same principle.

\subsection{OCC-based MMSE}
It is noted that, when $\Delta_{\text {cs}} = P/2$ with an even number
$P$ for simplicity, the two DMRS ports,
$x_{2p}$, $x_{2p+1}$ for the first port and  $x_{2p}$, $-x_{2p+1}$
for the second port with $p=0,1,\cdots,P/2-1$,
can be separated by the orthogonal cover code (OCC), i.e., $[1,1]$ and $[1,-1]$ .
In this subsection, the OCC-based MMSE channel estimation scheme is presented for the two DMRS ports.

Firstly, equation (\ref{eq 2.1}) can be rewritten as
\begin{align}\label{eq 2.8}
\begin{split}
\left\{
\begin{array}{*{20}l}
   y_{2p}^{r} / x_{2p} = h_{2p}^{1r} + h_{2p}^{2r} + \eta_{2p}^{r}/x_{2p},\\
   \\
   y_{2p+1}^{r} / x_{2p+1} = h_{2p+1}^{1r} - h_{2p+1}^{2r} + \eta_{2p+1}^{r}/x_{2p+1}, ~~p=0,1,\cdots, P/2-1.\\
\end{array}\right.
\end{split}
\end{align}
Note that, $x_p$ has transmitting power of 1, therefore, noise $\eta_{p}^{r}/x_{p}$ satisfies the Gaussian
distribution with mean 0 and $\sigma^2$.

Assume that channel frequency responses, $h_{2p}^{tr}$ and $h_{2p+1}^{tr}$ are equal, i.e., $h_{2p}^{tr}=h_{2p+1}^{tr}$. Then, equation (\ref{eq 2.8}) can be rewritten as
\begin{align}\label{eq 2.9}
\begin{split}
\left\{
\begin{array}{*{20}l}
   y_{2p}^{r} / x_{2p} = h_{2p}^{1r} + h_{2p}^{2r} + \eta_{2p}^{r}/x_{2p},\\
   \\
   y_{2p+1}^{r} / x_{2p+1} = h_{2p}^{1r} - h_{2p}^{2r} + \eta_{2p+1}^{r}/x_{2p+1}, ~~p=0,1,\cdots, P/2-1.\\
\end{array}\right.
\end{split}
\end{align}
Denote $\bar{y}_{p}^{r}= y_{p}^{r} / x_{p}$ and $\bar{\eta}_{p}^{r}= \eta_{p}^{r} / x_{p}$.
Then,
 each user
can perform the channel estimation  independently as
\begin{align}\label{eq 2.10}
\begin{split}
\left\{
\begin{array}{*{20}l}
     \hat{h}_{2p}^{1r} = \frac{\bar{y}_{2p}^{r}   + \bar{y}_{2p+1}^{r} }{2} = {h}_{2p}^{1r} +  \frac{\bar{\eta}_{2p}^{r}   + \bar{\eta}_{2p+1}^{r} }{2} ,\\
   \\
   \hat{h}_{2p}^{2r} = \frac{\bar{y}_{2p}^{r}   - \bar{y}_{2p+1}^{r} }{2} ={h}_{2p}^{2r} +
    \frac{\bar{\eta}_{2p}^{r}   - \bar{\eta}_{2p+1}^{r} }{2}, ~~p=0,1,\cdots, P/2-1.\\
\end{array}\right.
\end{split}
\end{align}

Then, the channel estimation can be improved by the MMSE criteria, which can be obtained as
\begin{equation}\label{eq 2.11}
\hat{\textbf{h}}_{tr}^{\textrm{OCC-MMSE}}=\textbf{R}_{tr} \left(\textbf{R}_{tr}  + \sigma^2 \textbf{I}_{P/2}\right)^{-1} \hat{\textbf{h}}_{tr}.
\end{equation}
where $\hat{\textbf{h}}_{tr} =[\hat{h}_{0}^{tr},\hat{h}_{2}^{tr},\cdots, \hat{h}_{P-2}^{tr}]^T$. $\textbf{R}_{{tr}}$ is the auto-covariances of $\textbf{h}_{tr}=[h_{0}^{tr},h_{2}^{tr},\cdots, h_{P-2}^{tr}]^{T}$.

It is worthwhile  noting that, the OCC-based MMSE is based the assumption  $h_{2p}^{tr}=h_{2p+1}^{tr}$,
which indicates a performance loss of channel estimation under highly frequency selective channels.
Especially, for the two DMRS ports as presented in Figure.~\ref{fig1}, the pilot symbols locate at
every two subcarriers.
Therefore, the
channel frequency responses at $m$-th and $(m+2)$-th subcarriers are required
to be equal in the classical OCC-based MMSE scheme.


\section{Proposed Schemes for Two-port DMRS}
In this section, we firstly present the F-MMSE scheme based on the full priori knowledge of the two ports for
channel estimation in NR.
When the two ports are assigned to different users and the full priori knowledge of two ports
is difficult to obtain for one user.
Then, the P-MMSE scheme is presented for two-port channel estimation, based the partial knowledge of only one port.
For simplicity, the
channel estimation of ${\textbf{h}}_{1r}$ is only presented in detail in this paper and the estimation of ${\textbf{h}}_{2r}$
can be done in the same way.

\subsection{F-MMSE Channel Estimation}

For the channel estimation model  in (\ref{eq 2.3}),  $\textbf{h}_{1r}$ is the random vector of parameters to be estimated. Then,   the F-MMSE channel estimation of $\textbf{h}_{1r}$ can be obtained by minimizing the
mean square error (MSE),  defined as
\begin{equation}\label{eq 3.1}
\textrm{MSE}=\mathbb{E}
\left\{\left\|\textbf{h}_{1r}-\hat{\textbf{h}}_{1r}\right\|_2^2\right\},
\end{equation}
where $\hat{\textbf{h}}_{1r}$ is an estimate of $\textbf{h}_{1r}$.
Subsequently, the F-MMSE channel estimation of $\textbf{h}_{1r}$ can be obtained by minimizing the MSE
\begin{equation}\label{eq 3.2}
\hat{\textbf{h}}_{1r}^{\textrm{F-MMSE}}=\textbf{R}_{\hat{\textbf{h}}\textbf{h}_{1r}}  \textbf{R}_{\hat{\textbf{h}}\hat{\textbf{h}}}^{-1} \hat{\textbf{h}},
\end{equation}
where $\textbf{R}_{\hat{\textbf{h}}\textbf{h}_{1r}}$  is the cross covariance matrix
between $\hat{\textbf{h}}$ and $\textbf{h}_{1r}$. Further,
\textcolor{black}{$\textbf{R}_{\hat{\textbf{h}}\hat{\textbf{h}}}$  is the auto covariance matrix
of $\hat{\textbf{h}}$. }
$\textbf{R}_{\hat{\textbf{h}}\textbf{h}_{1r}}$ and $\textbf{R}_{\hat{\textbf{h}}\hat{\textbf{h}}}^{-1}$ are obtained as
\begin{align}
\textbf{R}_{\hat{\textbf{h}}\textbf{h}_{1r}}
&= \mathbb{E}\left\{ \textbf{h}_{1r} \hat{\textbf{h}}^{\text \dag}\right\}
= \textbf{R}_{1r},\label{eq 3.3}\\
\textbf{R}_{\hat{\textbf{h}}\hat{\textbf{h}}}
&= \mathbb{E}\left\{ \hat{\textbf{h}} \hat{\textbf{h}}^{\text \dag} \right\}
= \textbf{R}_{1r} + \textbf{C}\textbf{R}_{2r}\textbf{C}^{\text \dag} + \sigma^2 \textbf{I}_{P},\label{eq 3.4}
\end{align}
respectively,
where $\textbf{I}_{P}$ is an identity matrix with dimension of $P$. $\textbf{R}_{{1r}}$ and $\textbf{R}_{{2r}}$ are auto-covariances of $\textbf{h}_{1r}$ and $\textbf{h}_{2r}$.  Therefore, based on the priori knowledge $\textbf{R}_{{1r}}$ and $\textbf{R}_{{2r}}$,
the F-MMSE can be obtained as
\begin{equation}\label{eq 3.4.0}
\hat{\textbf{h}}_{1r}^{\textrm{F-MMSE}}=\textbf{R}_{1r} \left(\textbf{R}_{1r} + \textbf{C}\textbf{R}_{2r}\textbf{C}^{\text \dag} + \sigma^2 \textbf{I}_{P}\right)^{-1} \hat{\textbf{h}}.
\end{equation}

As a remark, when only one of the two DMRS ports is used, i.e., no signal is transmitted via the second DMRS port,
the optimal MMSE channel estimation of $\textbf{h}_{1r}$
is
\begin{equation}\label{eq 3.4.1}
\hat{\textbf{h}}_{1r}^{\textrm{o}}=\textbf{R}_{1r} \left(\textbf{R}_{1r}  + \sigma^2 \textbf{I}_{P}\right)^{-1} \hat{\textbf{h}}.
\end{equation}
Note that although the above F-MMSE, i.e., MMSE, is straightforward, it helps us to develop a more practical estimator below.

\subsection{P-MMSE Channel Estimation}\label{p-mmse}
For (\ref{eq 3.4.0}), one has to get the priori knowledge of both the two DRMS ports, i.e., $\textbf{R}_{1r}$
and $\textbf{R}_{2r}$. However,
when the two DMRS ports are assigned to different users, one user performing the channel estimation does not know
the priori knowledge of another DMRS port and that whether another DMRS port is used or not.
In our proposed P-MMSE scheme, we replace $\textbf{R}_{2r}$ by $\textbf{R}_{1r}$ and the channel estimation is
written as
\begin{equation}\label{eq 3.5}
\hat{\textbf{h}}_{1r}^{\textrm{P-MMSE}}=\textbf{R}_{1r} \left(\textbf{R}_{1r} + \textbf{C}\textbf{R}_{1r}\textbf{C}^{\text \dag} + \sigma^2 \textbf{I}_{P}\right)^{-1} \hat{\textbf{h}}.
\end{equation}

To show the validity of the proposed P-MMSE scheme, the theoretical analysis is presented in the following.
Firstly, the frequency-domain estimation in (\ref{eq 3.5}) is converted into the time domain as
\begin{align}\label{eq 3.6}
\underline{\hat{\textbf{h}}}_{1r}^{\textrm{P-MMSE}} &= \textbf{F}\hat{\textbf{h}}_{1r}^{\textrm{P-MMSE}} = \textbf{F} \textbf{R}_{1r} \left(\textbf{R}_{1r} + \textbf{C}\textbf{R}_{1r}\textbf{C}^{\text \dag} + \sigma^2 \textbf{I}_{P}\right)^{-1}\textbf{F}^{\text \dag} \textbf{F} \hat{\textbf{h}}
\end{align}
Substituting (\ref{eq 2.4}) into (\ref{eq 3.6}), we have
\begin{align}\label{eq 3.7}
\underline{\hat{\textbf{h}}}_{1r}^{\textrm{P-MMSE}}
 &=\textbf{F} \textbf{R}_{1r} \left(\textbf{R}_{1r} + \textbf{C}\textbf{R}_{1r}\textbf{C}^{\text \dag} + \sigma^2 \textbf{I}_{P}\right)^{-1}\textbf{F}^{\text \dag}\left(\underline{{\textbf{h} }}_{1r} +\underline{{\textbf{h}}}_{2r} \left(\frac{P}{2}\right)  + \boldsymbol{\xi}_{r}\right) \nonumber \\
 &= \boldsymbol{\Phi}_{1r}^{\textrm{P-MMSE}} \left(\underline{{\textbf{h} }}_{1r} +\underline{{\textbf{h}}}_{2r}\left(\frac{P}{2}\right)  \right) + \boldsymbol{\Phi}_{1r}^{\textrm{P-MMSE}} \boldsymbol{\xi}_{r},
\end{align}
where $\boldsymbol{\Phi}_{1r}^{\textrm{P-MMSE}} = \textbf{F} \textbf{R}_{1r} \left(\textbf{R}_{1r} + \textbf{C}\textbf{R}_{1r}\textbf{C}^{\text \dag} + \sigma^2 \textbf{I}_{P}\right)^{-1}\textbf{F}^{\text \dag}$.
Note that the coefficient matrix of the proposed F-MMSE scheme is
$\boldsymbol{\Phi}_{1r}^{\textrm{F-MMSE}} = \textbf{F} \textbf{R}_{1r} \left(\textbf{R}_{1r} + \textbf{C}\textbf{R}_{2r}\textbf{C}^{\text \dag} + \sigma^2 \textbf{I}_{P}\right)^{-1}\textbf{F}^{\text \dag}$.

\begin{figure}[!t]
	\centering
	\includegraphics[scale=0.6]{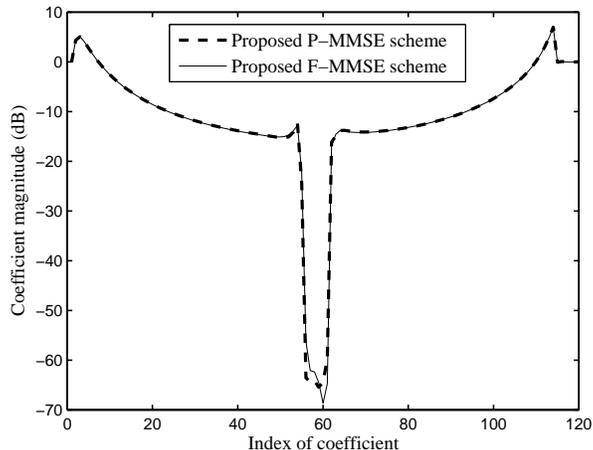}
	\caption{Magnitudes of the diagonal elements of the P-MMSE and F-MMSE coefficient matrices, i.e., $\boldsymbol{\Phi}_{1r}^{\textrm{P-MMSE}}$ and $\boldsymbol{\Phi}_{1r}^{\textrm{F-MMSE}}$, SNR = 30 dB.}
	\label{fig3}
\end{figure}

It is observed that from (\ref{eq 3.7}),
the estimation $\underline{\hat{\textbf{h}}}_{1r}^{\textrm{P-MMSE}} $ is a linear
combination of $\underline{{\textbf{h} }}_{1r} $
and $\underline{{\textbf{h}}}_{2r} \left(\frac{P}{2}\right)$ by the P-MMSE coefficient $\boldsymbol{\Phi}_{1r}^{\textrm{P-MMSE}}$. It can be easily seen that,
the values of  the channel estimation  mainly depend on
the diagonal coefficients of the matrix $\boldsymbol{\Phi}_{1r}^{\textrm{P-MMSE}}$.
As a result, Figure~\ref{fig3} depicts the magnitudes of the diagonal elements of the P-MMSE and F-MMSE coefficient matrices, in which $P=120$ and $M=2048$.
In addition, a random channel model is
adopted with 40 sample-spaced independent Rayleigh fading paths, which exhibit an exponential
power delay profile \cite{ref7}
as $\alpha(l) = e^{\beta l }$, where $l=0,1,\ldots,39$, $\beta = -0.0005$ for the first DMRS port
and $\beta = -0.05$ for the second DMRS port, i.e., $\beta = -0.0005$ is for the channels $\textbf{h}_{11}$ and
$\textbf{h}_{12}$, and $\beta = -0.05$ is for the channels $\textbf{h}_{21}$ and
$\textbf{h}_{22}$. The different  $\beta$ in channels lead to the different  channel covariance matrices.
From the simulation results, the  magnitude of the diagonal element of the P-MMSE coefficient matrix
is accordance to that of F-MMSE despite of the large difference of the attenuation factor $\beta$, exhibiting
insensitive to  channel covariance matrices. Furthermore,
it can be observed that, the P-MMSE coefficients have very small
values at the points around $\frac{P}{2}$, which indicates
that, for the channel estimation of $\underline{\textbf{h}}_{1r}$
the interference from $\underline{\textbf{h}}_{2r}\left(\frac{P}{2}\right)$ is greatly suppressed since channel coefficients of $\underline{\textbf{h}}_{2r}\left(\frac{P}{2}\right)$ have valid values
at round $\frac{P}{2}$ as you can see Figure~\ref{fig2}.
As a remark, it is worthwhile to note that, the P-MMSE coefficients have valid values at round both of  $0$ and $P$.
The reason is that when the subcarrier number of users is less than the system subcarrier, i.e., $P<M$,
the energy of channels coefficients is leaked at the points around $P$ due to the Gibbs phenomenon \cite{ref10}.

\subsection{MSE Analysis}

In this subsection, MSE of the proposed F-MMSE and P-MMSE estimators are analyzed.
Without loss of generality, only MSE with respect to ${\textbf{h}}_{1r}$ is considered.

Firstly,  MSE of the proposed F-MMSE can be obtained by
\begin{equation}\label{eq 3.8}
\textrm{MSE}_{F}=\textrm{Tr}\left\{ \mathbb{E}
\left[\left(\hat{\textbf{h}}_{1r}^{\textrm{F-MMSE}} - \textbf{h}_{1r}\right)\left(\hat{\textbf{h}}_{1r}^{\textrm{F-MMSE}} - \textbf{h}_{1r}\right)^{\dag} \right]\right\},
\end{equation}
where $\textrm{Tr}\{ \cdot\}$ denotes the trace operation.

It should be pointed out that, the channels  ${\textbf{h}}_{1r}$, ${\textbf{h}}_{2r}$, and noise $\boldsymbol{\eta}_{r}$ are independent. Therefore, it can be obtained
\begin{equation}\label{eq 3.9}
\mathbb{E}[{\textbf{h}}_{1r}{\textbf{h}}_{2r}^{\dag}]=0,
\end{equation}
\begin{equation}\label{eq 3.10}
\mathbb{E}[{\textbf{h}}_{1r}{\boldsymbol{\eta}_{r}}^{\dag}]=0,
\end{equation}
\begin{equation}\label{eq 3.11}
\mathbb{E}[{\textbf{h}}_{2r}{\boldsymbol{\eta}_{r}}^{\dag}]=0.
\end{equation}
For simplicity of presentation, we denote $\textbf{R}_{1r} \left(\textbf{R}_{1r} + \textbf{C}\textbf{R}_{2r}\textbf{C}^{\text \dag} + \sigma^2 \textbf{I}_{P}\right)^{-1}$ in
equation (\ref{eq 3.4.0}) as $\textbf{A}$.
Then, by substituting equations (\ref{eq 2.3}) and (\ref{eq 3.4.0}) into (\ref{eq 3.8}),
the MSE of the proposed F-MMSE estimator is easily obtained
\begin{equation}\label{eq 3.12}
\textrm{MSE}_{F}=\textrm{Tr}\left\{ \textbf{A} \textbf{R}_{1r}\textbf{A}^{\dag}  +  \textbf{A} \textbf{C} \textbf{R}_{2r} \textbf{C}^{\dag} \textbf{A}^{\dag} +  \textbf{A} \textbf{A}^{\dag}\sigma^2 -  \textbf{A} \textbf{R}_{1r} -  \textbf{R}_{1r}\textbf{A}^{\dag} +   \textbf{R}_{1r}           \right\}.
\end{equation}

Similarity, we denote $\textbf{R}_{1r} \left(\textbf{R}_{1r} + \textbf{C}\textbf{R}_{1r}\textbf{C}^{\text \dag} + \sigma^2 \textbf{I}_{P}\right)^{-1}$ in
equation (\ref{eq 3.5}) as $\textbf{B}$, and the MSE of the proposed P-MMSE estimator can be obtained
\begin{equation}\label{eq 3.13}
\textrm{MSE}_{P}=\textrm{Tr}\left\{ \textbf{B} \textbf{R}_{1r}\textbf{B}^{\dag}  +  \textbf{B} \textbf{C} \textbf{R}_{2r} \textbf{C}^{\dag} \textbf{B}^{\dag} +  \textbf{B} \textbf{B}^{\dag}\sigma^2 -  \textbf{B} \textbf{R}_{1r} -  \textbf{R}_{1r}\textbf{B}^{\dag} +   \textbf{R}_{1r}           \right\}.
\end{equation}

It can be observed that, MSE of F-MMSE and P-MMSE estimators depend on the MMSE coefficients, i.e.,
$\textbf{A}$  and $\textbf{B}$, respectively.
As presented in Subsection \ref{p-mmse}, the MMSE coefficient is insensitive to channel covariance mismatching.
As a result, the coefficients of $\textbf{A}$ is close to the coefficients of $\textbf{B}$, which indicates
the proposed P-MMSE estimator has a similar MSE performance with the F-MMSE estimator.

\section{Simulation Results}

In this section, numerical simulations are carried out to show the validity of the proposed  schemes
in a $2 \times 2$ multiple-antenna OFDM system with
$P=120$ and $M=2048$, in which the system sampling rate is 30.72MHz.   In simulations, quadrature phase shift keying (QPSK) modulation and no channel coding scheme are considered.
\textcolor{black}{
In addition, the length of cyclic prefix is 144 samples.}

\begin{figure}[!t]
	\centering
	\includegraphics[scale=0.6]{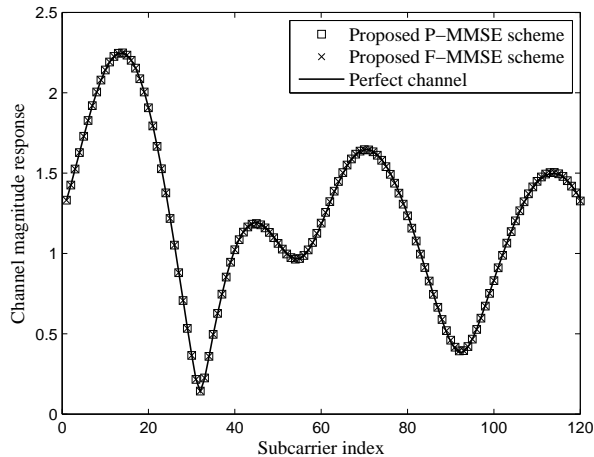}
	\caption{Channel magnitude responses of $\hat{\textbf{h}}_{11}^{\textrm{P-MMSE}}$, $\hat{\textbf{h}}_{11}^{\textrm{F-MMSE}}$  and  the true channel ${\textbf{h}}_{11}$, SNR = 30 dB.}
	\label{fig4}
\end{figure}

Figure~\ref{fig4} shows the channel magnitude responses of
the proposed P-MMSE and F-MMSE schemes to evaluate the validity of channel estimation, in which
the channel models in Subsection 3.2 are adopted and
signal-to-noise ratio (SNR) is set to $30$ dB to make the noise small enough.
It is easily observed that, the channel estimations of the proposed P-MMSE and F-MMSE schemes are
accordance to the perfect
channel ${\textbf{h}}_{11}$, without significant interference.
Note that, the P-MMSE scheme approaches to the same performance of the F-MMSE scheme with a little priori knowledge.

\begin{figure}[!t]
	\centering
	\includegraphics[scale=0.6]{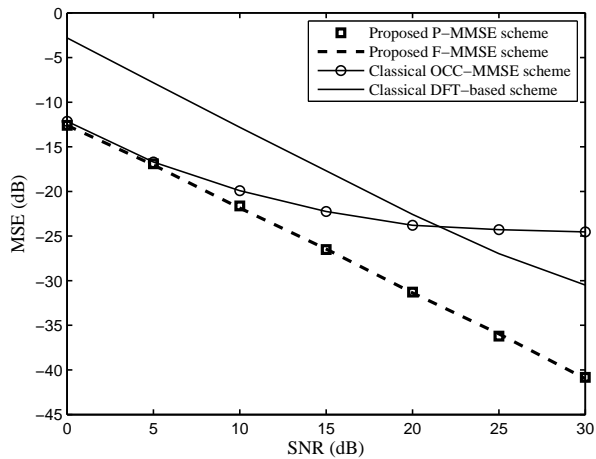}
	\caption{MSE comparison of the proposed P-MMSE scheme, the proposed F-MMSE scheme, and the classical  DFT-based scheme under channels with exponential power delay profiles.}
	\label{fig5}
\end{figure}

\begin{figure}[!t]
	\centering
	\includegraphics[scale=0.6]{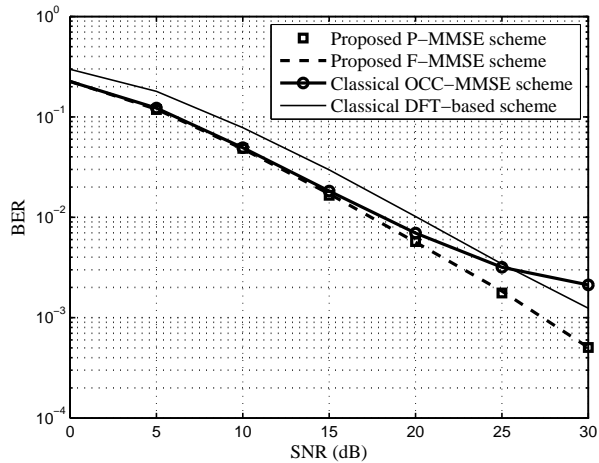}
	\caption{BER comparison of the proposed P-MMSE scheme, the proposed F-MMSE scheme,  and the classical  DFT-based scheme under channels with exponential power delay profiles.}
	\label{fig6}
\end{figure}

Figure~\ref{fig5} and Figure~\ref{fig6} depict the  mean square error (MSE) and bit error ratio (BER) performances
of the proposed P-MMSE and F-MMSE schemes, respectively. In simulations,
 the channel models in Subsection 3.2 are adopted.
For comparison, we also give the performances of the classical DFT-based scheme
and the OCC-based MMSE scheme. From simulation results, the proposed P-MMSE and F-MMSE schemes exhibit
large gains in terms of MSE and BER compared with the classical DFT-based scheme, since the classical DFT-based scheme
does not require the priori knowledge of channel covariance matrix.
It is also observed that, compared with the proposed P-MMSE and F-MMSE schemes, the classical OCC-based MMSE can achieve similar MSE and BER performances at low
SNR region and suffers from obvious loss at high SNR region. The reason is  that
the classical OCC-based MMSE is based on the assumption that the channel frequency responses at adjacent subcarrier are equal, which indicates performance loss under high frequency selective channels. Note that,
\textcolor{black}{as depicted in
Figure~\ref{fig1},} pilot symbols of the two DMRS ports locate at every two subcarriers, therefore, the
channel frequency responses at $m$-th and $(m+2)$-th subcarriers are required to be equal in the classical OCC-based MMSE scheme. On the other hand,
the proposed F-MMSE scheme can achieve the optimal channel estimation performance, with the full
priori knowledge of both DMRS ports, i.e., $\textbf{R}_{{1r}}$ and $\textbf{R}_{{2r}}$.
When the two DMRS ports are assigned to two different users, the full
priori knowledge is not easy to obtain for one user. The proposed P-MMSE only requires
the priori knowledge of one port and can
achieve the similar performance compared with the proposed F-MMSE, despite the large difference of the attenuation factors of exponential power delay profiles.


\begin{figure}[!t]
	\centering
	\includegraphics[scale=0.6]{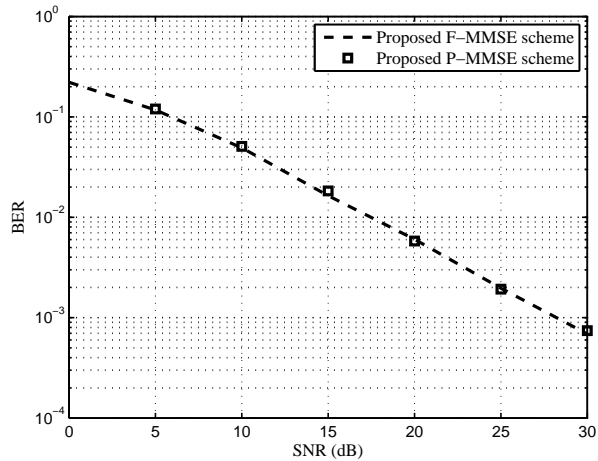}
	\caption{BER comparison of the proposed P-MMSE scheme and the proposed F-MMSE scheme, under the TDL channels.}
	\label{fig7}
\end{figure}

To evaluate the proposed schemes under more practical channel models, tapped delay line (TDL)
models \cite{ref11} released by 3GPP are also used in simulations.
Note that, the two DMRS ports are assigned to different channel models. Specifically,
the first DMRS port is with TDL-A  and the second DMRS port is with TDL-C, in which TDL-A  and TDL-C
exhibit different delay spreads and power delay profiles \cite{ref11}.
Figure~\ref{fig7} shows BER comparison of the proposed P-MMSE scheme and F-MMSE scheme under the TDL channels.
From simulation results, the proposed P-MMSE
achieves the similar BER performance compared with the optimal F-MMSE, which validates the effectiveness of the proposed
P-MMSE scheme.

\begin{figure}[!t]
	\centering
	\includegraphics[scale=0.6]{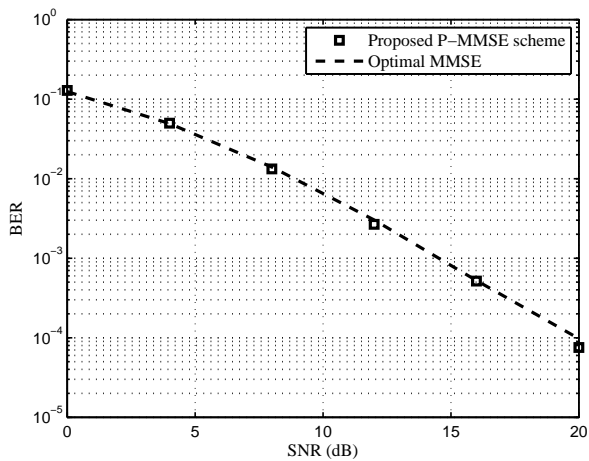}
	\caption{BER performance of the proposed P-MMSE scheme  under the TDL channels when only one of the
two DMRS ports is used.}
	\label{fig8}
\end{figure}

When only one of the two DMRS ports is used, the $2 \times 2$ multiple-antenna OFDM system becomes
a $1 \times 2$ multiple-antenna system. In this case, the optimal MMSE channel estimation will be
equation (\ref{eq 3.4.1}). However, for one user with the first DMRS port, it is difficult to
obtain the priori knowledge on whether another DMRS port is used or not.
Figure~\ref{fig8} shows the
BER performance of the proposed P-MMSE scheme  under the TDL channels when only one of the
two DMRS ports is used. For comparison, the performance of the optimal MMSE as (\ref{eq 3.4.1})
is also given. From simulation results, the proposed P-MMSE
achieves a similar BER performance compared with the optimal MMSE.

\begin{figure}[!t]
	\centering
	\includegraphics[scale=0.6]{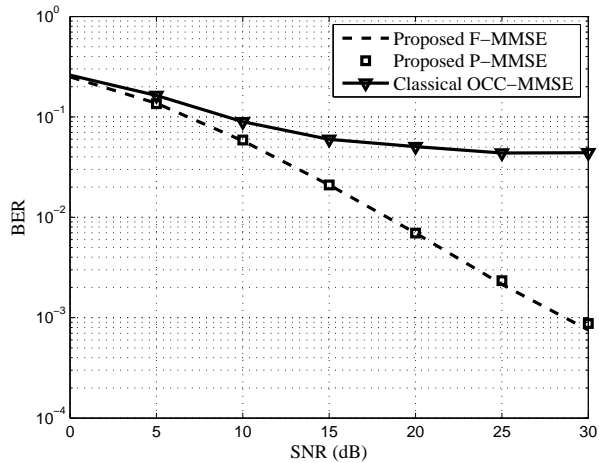}
	\caption{BER performance of the proposed P-MMSE scheme  under TDL-C and the channel with equal-power
 delay profiles.}
	\label{fig9}
\end{figure}

Then, the first DMRS port is with the TDL-C channel and the second DMRS port is with the channel with
equal-power power delay profiles and maximum channel delay spread $9.65860 \mu s$.
Figure~\ref{fig9} shows BER comparison of the proposed P-MMSE scheme and F-MMSE scheme under the TDL channel
and the channel with equal-power  delay profiles \cite{addzhang}.
From simulation results, the proposed P-MMSE
achieves the similar BER performance compared with the optimal F-MMSE, which validates the effectiveness of the proposed
P-MMSE scheme.
Besides,
the classical OCC-based MMSE  suffers from obvious performance loss. The reason is  that
the classical OCC-based MMSE is based on the assumption that the channel frequency responses at adjacent subcarrier are equal, which is not satisfied well under the high frequency selective channels.

\section{Conclusion}
In this paper, we firstly proposed the F-MMSE scheme to achieve the channel estimation of
two-port DMRS in NR, which requires the full priori knowledge of the channel covariance matrix of two ports.
Then, we presented the P-MMSE scheme
with the priori knowledge of only one DMRS port.
Finally, theoretical analysis and numerical simulations showed that the proposed two schemes achieve
satisfactory channel estimation performance.
Also, the proposed P-MMSE scheme
is insensitive to  the difference of channel covariance matrices
of the two DMRS ports, and can achieve the similar performance compared with the  F-MMSE scheme. Both of the proposed P-MMSE and F-MMSE schemes have large gains in terms of MSE and BER compared with the classical DFT-based scheme.

%
%


\begin{thebibliography}{99}

\bibitem{ref2.0}
W. Ma, C. Qi, and G. Y. Li, ``High-resolution channel estimation for
frequency-selective mmWave massive MIMO systems," \textit{IEEE Trans. Wireless Commun.}, vol. 19, no. 5, pp. 3517-3529, May 2020.

\bibitem{ref0}
J. Fan, Q. Yin, and W. Wang, ``Pilot-aided channel estimation for CDD-OFDM systems," \textit{Sci. China Inf. Sci.}, vol. 53, no. 2, pp. 379-389, Feb 2010.



\bibitem{ref1}
A. Ghosh, R. Ratasuk, B. Mondal, N. Mangalvedhe,
and T. Thomas, ``LTE-Advanced: next-generation wireless broadband technology," \textit{IEEE Wireless Commun.}, vol. 17, no. 3, pp. 10-22, Jun. 2010.



\bibitem{ref00}
Y. Li, X. Xu, D. Zhang, Z. Zhang, and K. Long, ``Optimal pilots design for frequency offsets and channel estimation in OFDM modulated single frequency networks," \textit{Sci. China Inf. Sci.}, vol. 57, Apr. 2014.




\bibitem{ref2}
M. El-Absi, S. Galih, M. Hoffmann, M. El-Hadidy, and T. Kaiser, ``Antenna selection for reliable MIMO-OFDM interference alignment systems: measurement-based evaluation," \textit{IEEE Trans. Veh. Technol.}, vol. 65, no. 5, pp. 2965-2977, May. 2016.

\bibitem{ref2.11}
S. Nagaraj, ``Pre-DFT combining for coded OFDM," \textit{IEEE Trans. Veh. Technol.}, vol. 58, no. 9, pp. 5305-5309, Nov. 2009.


\bibitem{ref3}
3GPP, ``NR; Physical channels and modulation (Release 15)," v15.1.0, TS 38.211, 2018. [Online]. Available: http://www.3gpp.org/ftp//Specs/archive/38\_series/38.211/38211-f10.zip


\bibitem{ref3.0}
C. B. Barneto, T. Riihonen, M. Turunen, L. Anttila, M. Fleischer, K. Stadius, J. Ryyn\"{a}nen, and M. Valkama, ``Full-duplex OFDM radar with LTE and 5G NR waveforms: challenges, solutions, and measurements," \textit{IEEE Trans. Microw. Theory Techn.}, vol. 67, no. 10, pp. 4042-4054, Aug. 2019.

\bibitem{ref4}
M. Jiang, G. Yue, N. Prasad, and S. Rangarajan. ``Enhanced DFT-based channel estimation for LTE
uplink," in \textit{Proc. IEEE VTC Spring}, May 2012.


\bibitem{ref5}
X. Li, X. Jing, S. Sun, H. Huang, and Y. Lu. ``An improved DFT-based channel estimation method for OFDM system," in \textit{Proc. IET CCT}, Nov. 2013,  pp. 550-554.

\bibitem{ref6}
J. Zhang, and L. Huang. ``An improved DFT-based channel estimation algorithm for MIMO-OFDM systems," in \textit{Proc. CECNET}, Apr. 2011, pp. 3929-3932.

\bibitem{Sesia}
S. Sesia, I. Toufik, and M. Baker, \textit{LTE-The UMTS Long Term Evolution: From Theory to Practice}. 
West Sussex, U.K.: John Wiley \& Sons. Ltd., 2011.


\bibitem{ye}
Y. Li,  L. J. Cimini, Jr., and N. R. Sollenberger, ``Robust channel estimation for OFDM systems
with rapid dispersive fading channels," \textit{IEEE Trans. Commun.}, vol. 46, no. 7, pp. 902-915, Jul. 1998.

\bibitem{addzhang}
Y. Zhang, D. Wang, J. Wang, and X. You, ``Channel estimation for massive MIMO-OFDM
systems by tracking the joint angle-delay subspace," \textit{IEEE Access}, vol. 4, pp. 10166-10179, 2016.



\bibitem{ref9.1}
P. Liu, S. Jin, T. Jiang, Q. Zhang, and M. Matthaiou, ``Pilot power allocation through user grouping
in multi-cell massive MIMO systems," \textit{IEEE Trans. Commun.}, vol. 65, no. 4, pp. 1561-1574, Apr. 2017.

\bibitem{ref9.11}
D. Kong, X.-G. Xia, and T. Jiang, ``An alamouti coded CP-FBMC-MIMO system with two transmit antennas," \textit{Sci. China Inf. Sci.}, vol. 58, Oct. 2015.

\bibitem{ref7}
D. Kong, D. Qu, and T. Jiang, ``Time domain channel estimation for
OQAM-OFDM systems: algorithms and
performance bounds," \textit{IEEE Trans. Signal Process.}, vol. 68, no. 2, pp. 322-330, Jan. 2014.






\bibitem{ref10}
X. Hou, Z. Zhang, and H. Kayama, ``DMRS design and channel estimation for LTE-advanced MIMO uplink," in \textit{Proc. IEEE VTC Fall},  Sep. 2009.

\bibitem{ref11}
3GPP, ``Technical specification group radio access network; study on channel model for frequencies from 0.5 to 100 GHz (Release 16),'' v16.1.0, TS 38.901, 2019. [Online]. Available: https://www.3gpp.org/ftp/Specs/archive/38\_series/38.901/38901-g10.zip


\end{thebibliography}
\end{document}